\documentclass[aps,pre,amsmath,amssymb]{revtex4-2}

\usepackage{graphicx} 
\usepackage{xcolor}
\usepackage[caption = false]{subfig}
\usepackage{graphicx}
\usepackage{booktabs}
\usepackage{natbib}
\usepackage{amsthm}
\usepackage{amsmath}
\usepackage{amssymb}
\usepackage{geometry}
\usepackage{algpseudocode}
\usepackage{multirow}

\usepackage{lineno}
\usepackage{hyperref}

\usepackage[xcolor=dvipdf]{changes}
\definechangesauthor[name={MC}, color=orange]{MC}
\definechangesauthor[name={PP}, color=red]{PP}

\newcommand{\beq}{\begin{equation}}
\newcommand{\eeq}{\end{equation}}

\begin{document}


\title{Criticality of a Stochastic Dense Associative Memory Model with Exponential Interaction Function}


\author{Marco Cafiso}
\email[]{Corresponding author: marco.cafiso@cnr.it}
\affiliation{Department of Physics `E. Fermi', University of Pisa, Largo Bruno Pontecorvo 3, I-56127, Pisa, Italy}
\affiliation{Institute of Information Science and Technologies `A. Faedo', ISTI-CNR, Via G. Moruzzi 1, I-56124, Pisa, Italy}

\author{Paolo Paradisi
}
\email[]{Corresponding author: paolo.paradisi@cnr.it}
\affiliation{
Institute of Information Science and Technologies ‘A. Faedo’, ISTI-CNR, Via G. Moruzzi 1, I-56124, Pisa, Italy}
\affiliation{
BCAM-Basque Center for Applied Mathematics, Alameda de Mazarredo 14, E-48009, Bilbao, BASQUE COUNTRY, Spain}


\date{\today}



\begin{abstract}
{\bf Abstract:}\\
The Hopfield network (HN) is a classical model of associative memory with stored patterns encoded as minima of an energy function shaped by a Hebbian learning rule.
Dense Associative Memory (DAM) models introduce $n$-body interactions among neurons with $n>2$ and, more recently, also exponential interaction functions, which significantly improve the network's storing capacity. 
While the emergence of phase transitions in 
HN and DAM were extensively studied, 
the investigation of exponential DAM is still in its early stages. Further, an equilibrium thermodynamical condition is typically assumed, while out-of-equilibrium dynamics are not considered.\\
Here, we study the temporal dynamics of a stochastic exponential DAM (SEDAM) with a multiplicative salt-and-pepper noise and trained on the MNIST dataset. 
While taking the noise probability $p$ as control parameter, the time-averaged overlap ${\overline Q}$ and the diffusion scaling $H$ 
are taken as order parameters, being $H$ related to the network's time correlation features.
The MNIST-based SEDAM is also compared with a SEDAM trained on standard Rademacher patterns and with a stochastic HN (SHN).
%
%

\noindent
We found
the emergence of a phase transition in both ${\overline Q}$ and $H$, with the critical noise level $p^c$ decreasing as the load $K$ increases.
For each load $K$, ${\overline Q}$ highlights a transition between a sub- and a super-critical regime, both with short-time correlated dynamics.
Conversely, in the critical regime of the MNIST-based SEDAM the network displays a long-time correlated dynamics with highly persistent temporal memory marked by the high value $H \simeq 1.25$.
Similar behaviors are observed for both models trained with Rademacher patterns, but with a slightly higher temporal memory index $H \simeq 1.5$.
\end{abstract}



\maketitle

\section{Introduction}
\label{intro}

\noindent
Associative memory models are well-known neural network models that apply a Hebbian learning rule \cite{hebb_1949,brown_ARN1990_HebbianReview}, which allows for storing a set of input patterns in the network structure through a proper definition of the link weights among neurons.
The first associative memory models were proposed and investigated by Little and others since the 1970' \cite{little_mb1974,little_bb1975,little_mb1978,amari_bc1977,amari_bc1978,pastur_tmp1978} and, among others, the classical Hopfield network (HN) model \cite{hopfield_pnas1982} and other Hopfield-type models are nowadays the most studied ones.
The classical HN is a discrete model in both time and state space, characterized by dichotomous McCulloch and Pitts neurons in which the subthreshold dynamics \footnote{
The term ``subthreshold'', or ``under-threshold'', refers to the dynamical rule that allows the neuron potential to reach the firing threshold, after which this same potential is reset to ``no firing'' condition, which is here given by the value $-1$.
}
are described by a discrete integration of neuron states multiplied by the strength of links between them \footnote{
McCulloch and Pitts neurons usually assume values in the set $\{- 1,+1\}$ or in $\{0,1\}$, but whatever couple of values $\{a,b\}$ could be used.
}.
McCulloch and Pitts neurons admit state values $S \in \{-1 , +1 \}$, which define ``no firing'' and ``firing'' states, respectively\footnote{
The value $\xi = 0$ is often used to define the ``no firing'' state.
}
%
The Hebbian learning rule can be synthesised into the phrase:``neurons that fire together, wire together'' \cite{hebb_1949, Shatz1992DevelopingBrain}. 
In a nutshell, given $K$ 
patterns $\mbox{\boldmath$\xi$}_\mu = [\xi_{\mu}^1, \xi_{\mu}^2, ...., \xi_{\mu}^N], \xi_{\mu}^i \in \{-1,1\}$, $\mu=1,...,K$, 
the HN model applies the Hebbian rule by simply exploiting the following formula for the weight, or adjacency, matrix:
\begin{equation}   
W^{ij} = \frac1N \sum_{\mu =1}^K \xi_\mu^i \xi_\mu^j
\label{hebb_rule_2}
\end{equation}
Importantly, the Hebbian rule shapes an energy landscape where each pattern corresponds to an energy well, that is, a fixed point in the network dynamics with a local minimum of energy, being the energy function defined by:
\beq
E[\mbox{\boldmath$S$}] = 
- \sum_{
\substack{i,j=1 \\ i\ne j}
}^N W^{ij} S^i S^j
\label{energy_hn}
\eeq
%
%
with $S^i \in \{-1, +1\}$.
The dynamical equations governing the evolution of neurons in the HN are then given by \cite{hopfield_pnas1982,krotov_2016modernHopf_first}:
\beq
\left\{\begin{array}{c}
S^i_t = 1\ \ {\rm if } \ \ I^i_{t-1} > U  \\
\ \\
S^i_t = - 1 \ \ {\rm if } \ \ I^i_{t-1} \le U
\end{array}
\right.
\label{equations_hn}
\eeq
being $I^i_{t-1} = \sum_{j \ne i} W^{ij} S^j_{t-1}$ 
and $S_t^i$ the state of the $i$-th neuron and time $t$ \footnote{
In the original Hopfield proposal \cite{hopfield_pnas1982} and in many successive studies, the update dynamics are taken asynchronously, implying that each neuron randomly updates its state in a given time window. On the contrary, we here consider the case of synchronous update given in Eq. (\ref{equations_hn}). 
}. 
The firing threshold $U$ is here taken homogeneous throughout the network.
%
The update dynamics of Eq. (\ref{equations_hn})
lead to a path within the energy landscape:
\beq
E_t = E[\mbox{\boldmath$S$}_t] = 
- \sum_{
\substack{i,j=1 \\ i\ne j}
}^N W^{ij} S_t^i S_t^j
\label{energy_hn_time}
\eeq
and Hopfield, in his milestone paper \cite{hopfield_pnas1982}, proved that this pattern-induced energy landscape allowed the neural network to retrieve the stored pattern $\mbox{\boldmath$\xi$}_\mu$ when a distorted version $\mbox{\boldmath$S$}_{t=0} = \widetilde{\mbox{\boldmath$\xi$}}_\mu$,
e.g., a blurred image, is given as an initial condition to the network. This means that $E_t$ in Eq. (\ref{energy_hn_time}) evolves towards a local minimum, corresponding to a fixed point of the dynamics.
%
Clearly, every fixed point has a basin of attraction, which constrains the extent of distortion, and the retrieval is reliable when the number $K$ of stored patterns, also named {\it network load}, does not exceed a critical value \cite{hopfield_pnas1982,amit_pra1985,forrest_jpa1988}, which then defines the network {\it storage capacity}.
%
%
%
The {\it retrieval phase} is then defined by the parametric region where the network preserves the associative memory property, that is, the ability to retrieve the right pattern starting from an initial distorted pattern \footnote{
Clearly, the distorted pattern 
must belong to the attraction basin of the stored pattern itself. 
}. 
When the network load exceeds a critical value, there is a sharp drop in the retrieval ability that marks the transition to a ``forgetting'' phase \cite{hopfield_pnas1982}.
Notably, these emerging transitions between phases with different memory retrieval capacities have earned the Hopfield model a prominent place in the field of critical phenomena since the 1980s (see, e.g., \cite{amit_pra1985,gardner_epl1987,amit_ap1987,amit_pra1987}).


%
%
%
%

\noindent
To enlarge the capacity of classical HN models, many authors have proposed and studied generalized network models, named Dense Associative Memory (DAM) models, Modern Hopfield Networks (MHNs)
or, more generally, dense neural networks.
In these models a generalized n-body Hebbian rule replaces the 2-body interaction function of Eqs. (\ref{hebb_rule_2}) and (\ref{energy_hn}). More recently, an exponential interaction function, which will be introduced in Section \ref{sec:mhn}, has also been proposed and investigated \cite{demircigil2017modernHopfExp,albanese_pa2026}.
%
Similarly to the criticality of the classical HN model, the investigation of storage capacity in terms of both first- and second-order (critical) phase transitions in $n$-body DAM models is at the heart of a substantial body of research and, following significant advances in Artificial Intelligence (AI) studies,
has attracted growing interest over the past decade
\cite{amit_prl1985,amit_ap1987,amit_pra1987,gardner_epl1987,gardner_jpamg1987,bovier_atmp2001,krotov_2016modernHopf_first,krotov_nc2018,krotov_pnas2019,krotov_2021largeassociativememoryproblem,bao_nc2022,krotov_nrp2023,lucibello_prl2024,theriault_spp2024,theriault_nn2025,aguilera_nc2025,albanese_pa2026}.

\vspace{.15cm}

\noindent
Noteworthy, while many works and results are available on both the classical HN and the $n$-body DAM models,
there are still very few studies dedicated to the critical behavior emerging in the DAM model with the exponential interaction function of Demircigil et al.  \cite{demircigil2017modernHopfExp,ramsauer2021hopfieldnetworksneed,lucibello_prl2024} or other similar Hamiltonian functions based on exponential interaction \cite{albanese_pa2026}.
%
%
%
Further, this extensive body of research is based on the assumption of thermal equilibrium with a heat bath (canonical ensemble) and, by construction, does not capture out-of-equilibrium dynamics.\\
%
To explore these out-of-equilibrium dynamics, in this work we introduce a stochastic version of the exponential DAM (SEDAM) model, i.e., with the exponential energy function introduced by Demircigil et al. \cite{demircigil2017modernHopfExp} and we focus our study on its temporal behavior. The noise component is driven by a probability $p$, which will be introduced in more detail in Section \ref{sec:mhn} and plays the role of a control parameter.
Then, we characterize the emergence of phase transitions in the context of order parameters derived directly from the temporal evolution of the neural network, as determined by its dynamics. To this aim, we introduce a well-defined time-dependent overlap parameter, for which one-time (mean) and two-time statistical indices are computed. In particular, the two-time index, which will be introduced in Section \ref{sec:order-param} and Appendix \ref{app:dfa} is based on evaluating the scaling of a diffusion process generated by the time-dependent overlap parameter, an approach extensively applied to characterize the emergence of long-range correlations in complex systems \cite{kantelhardt_pa2001,mallick_tac2021,wang_ieeesj2022,kantelhardt_pa2002,paradisi_npg12,paradisi_springer2017}.\\
Thus, this work lies within a body of studies concerning the links between the emergence of phase transitions—particularly critical transitions—on the one hand, and the emergence of complexity on the other \cite{contoyiannis_pla2000,contoyiannis_prl2002,Contoyiannis_pre2007,allegrini_csf13,allegrini_pre15,tagliazucchi-chialvo_fn2016,turalska-grigolini_pre2011,beig-grigolini_pre15,zare-grigolini_csf2013,grigolini_chapt2014,mafahim-grigolini_njp2015}, which is typically associated with the onset of long-range temporal correlations.\\
The paper is organized as follows. 
In the next Section \ref{sec:mhn} we present the stochastic exponential DAM (SEDAM) model and, analogously, the stochastic version of the classical 2-body HN model (SHN), which we compare with each other.
In Section \ref{sec:order-param}, we define the order parameters that are here computed to characterize the emergence of a critical region. In Section \ref{sec:results} we give the results of numerical simulations and statistical data analyses, in Section \ref{sec:discuss} we discuss our findings and, finally, in Section \ref{sec:concl} we sketch some conclusions.

\section{The Dense Associative Memory network model}
\label{sec:mhn}
%

\noindent
In DAM models, the energy function has the
general expression \cite{krotov_2016modernHopf_first}:
\begin{equation}
E = -\sum_{\mu=1}^K F(\mbox{\boldmath$\xi$}_\mu^T \mbox{\boldmath$S$}_t)
\label{energy_krotov}
\end{equation}
where \(\mbox{\boldmath$\xi$}_\mu\) is the vector of the $\mu$-th stored pattern, $K$ the network load,
\(\mbox{\boldmath$S$}_t\) the $N$-dimensional vector of neuron states at time $t$, \(\mbox{\boldmath$\xi$}_\mu^T \mbox{\boldmath$S$}_t\) the scalar product between \(\mbox{\boldmath$\xi$}_\mu\) and \(\mbox{\boldmath$S$}_t\), and $F(\cdot)$ an interaction function. 
In Refs. \cite{krotov_2016modernHopf_first} and \cite{gardner_jpamg1987}, $F(z)$ is a polynomial function ($F(z) = z^n$), while in \cite{demircigil2017modernHopfExp}, it is an exponential function ($F(z) = \exp(z)$).
The classical HN is recovered for $n=2$, that is:
\begin{eqnarray}
E = -\sum_{\mu=1}^K 
[\mbox{\boldmath$\xi$}_\mu^T \mbox{\boldmath$S$}_t]^2 
&=& -\sum_{\mu=1}^K 
\mbox{\boldmath$S$}_t^T
[\mbox{\boldmath$\xi$}_\mu \mbox{\boldmath$\xi$}_\mu^T ]
\mbox{\boldmath$S$}_t 
=
\nonumber \\
&=& 
- \mbox{\boldmath$S$}^T  \mbox{\boldmath$W$}
\mbox{\boldmath$S$}_t \ ,
\label{classical_hn}
\end{eqnarray}
thus getting the well-known Hebbian rule given by Hopfield in his seminal paper \cite{hopfield_pnas1982}:
\begin{equation}
\mbox{\boldmath$W$} =  \sum_{\mu=1}^K 
\mbox{\boldmath$\xi$}_\mu \mbox{\boldmath$\xi$}_\mu^T\ .
\label{hebb_rule}
\end{equation}

\noindent
Following Krotov and Hopfield \cite{krotov_2016modernHopf_first}, we get the following dynamical update rule:
\begin{equation}
\mbox{\boldmath$S$}^i_{t} = \text{sgn} \Big[-E(\mbox{\boldmath$S$}_{t-1}^{(i+)}) + E(\mbox{\boldmath$S$}_{t-1}^{(i-)}) \Big]\ ,
\label{krotov_dynamics}
\end{equation}
where $i = 1,...,N$, \(\mbox{\boldmath$S$}_{t-1}^{(i+)}=\mbox{\boldmath$S$}_{t-1}(\mbox{\boldmath$S$}^i_{t-1}=+1)\) and \(\mbox{\boldmath$S$}_{t-1}^{(i-)}=\mbox{\boldmath$S$}_{t-1}(\mbox{\boldmath$S$}^i_{t-1}=-1)\).

\noindent
Substituting the exponential interaction function $F(z) = \exp(z)$ the previous equation becomes \cite{demircigil2017modernHopfExp}:
\begin{eqnarray}
\mbox{\boldmath$S$}^i_{t} = \text{sgn} \Bigg[ 
&-& \sum_{\mu=1}^K \exp(\mbox{\boldmath$\xi$}_\mu^T \mbox{\boldmath$S$}_{t-1}^{(i+)}) + 
\nonumber \\ 
&+& \sum_{\mu=1}^K \exp(\mbox{\boldmath$\xi$}_\mu^T \mbox{\boldmath$S$}_{t-1}^{(i-)}) \Bigg]
\label{modern_hop_exp}
\end{eqnarray}
Then, we define our SEDAM model with exponential interaction function by simply multiplying the previous expression for a dichotomous random process, thus getting the following dynamical equation in discrete time:
\begin{eqnarray}
\mbox{\boldmath$S$}^i_{t} = \epsilon^i_{t} 
\, \text{sgn} \Bigg[
&-& \sum_{\mu=1}^K \exp\left(\mbox{\boldmath$\xi$}_\mu^T \mbox{\boldmath$S$}_{t-1}^{(i+)}\right) + 
\nonumber \\ 
&+&
\sum_{\mu=1}^K \exp\left(\mbox{\boldmath$\xi$}_\mu^T \mbox{\boldmath$S$}_{t-1}^{(i-)}\right) \Bigg]\ ,
\label{modern_hop_exp_noise}
\end{eqnarray}
where, for each neuron $i =1,...,N$, $\epsilon^i_{t}$ is a binary random process in the time $t$ that takes values in $\{-1, 1\}$ with probability $P(\epsilon^i_{t} = -1) = p$ and $P(\epsilon^i_{t} = +1) = 1-p$. The random process $\epsilon^i_{t}$ has no correlations in time and among neurons and  it is also denoted as {\it salt-and-pepper} noise. As anticipated in the Introduction, the noise probability $p$ plays the role of the control parameter.
For sufficiently high $p$, the noise allows the system to make random jumps between different wells of the energy function, thus broadening the regions explored by the system and greatly enriching the network behavior.
Conversely, the condition of exact pattern retrieval is possible, for $K$ small enough, in the absence of noise ($p=0$), as the system rapidly falls into the nearest energy well.\\
%
%
%
Increasing $p$ from $0$ to $0.5$ is associated with an increase of disorder in the network model, with $p=0.5$ giving the maximum disorder. It is worth noting that the range of values $p=[0.5,1]$ is symmetric to the range $[0,0.5]$. In fact, at equal $K$, the value $1-p$ gives the same results as $p$, but with opposite signs in the values of neuron states.

\noindent
For comparison, we also consider the stochastic 2-body HN (SHN) model, which is given by considering $F(z) = z^2$ in Eq. (\ref{energy_krotov}), thus getting the following expression of the dynamics:
\begin{eqnarray}
\mbox{\boldmath$S$}^i_{t} = \epsilon^i_{t} 
\, \text{sgn} \Bigg[
&-& \sum_{\mu=1}^K \left(\mbox{\boldmath$\xi$}_\mu^T \mbox{\boldmath$S$}_{t-1}^{(i+)}\right)^2 + 
\nonumber \\ 
&+& \sum_{\mu=1}^K \left(\mbox{\boldmath$\xi$}_\mu^T \mbox{\boldmath$S$}_{t-1}^{(i-)}\right)^2 \Bigg]\ ,
\label{classical_hop_noise}
\end{eqnarray}
where the multiplicative noise $\epsilon^i_{t}$ is defined as in the SEDAM model of Eq. (\ref{modern_hop_exp_noise}).

\section{Overlap as order parameter}
\label{sec:order-param}
%

\noindent
To study the criticality of the SEDAM model by changing the number $K$ of stored patterns and the noise probability $p$, we refer to an overlap parameter as a measure of the model's ability to retrieve the correct pattern.
Without noise, we always have $\epsilon^i_t=1, \forall t,i$
in Eq. (\ref{modern_hop_exp_noise}) and, when $K$ is not too large, the neural network rapidly reaches an exact steady (equilibrium) state, given by the fixed point that is nearest to the initial distorted pattern.
Thus, the network finds the correct pattern when a blurred pattern is given as initial input.
In this zero-noise and low $K$ case, most authors study the following overlap parameter:
$$
Q^\mu_{\infty} = \lim_{t\rightarrow \infty} \frac{1}{N} \sum_{i=1}^N S^i_t\xi^i_\mu
$$
or other similar ones such as, e.g., the maximum over $\mu$ of $Q^\mu_{\infty}$.
The single-pattern overlap parameter $Q^\mu_{\infty}$ measures the correlation between the expected pattern $\mbox{\boldmath$\xi$}_\mu$ and the final output of the model, being the initial condition ${\bf S}_0$ a blurred version of the pattern $\mbox{\boldmath$\xi$}_\mu$, and a value $Q^\mu_{\infty} = 1$ is expected.
%

\noindent
Here, we initialize our simulations with a distorted version of the first stored pattern, $\widetilde{\mbox{\boldmath$\xi$}}_0$, so the overlap parameter is calculated concerning the first pattern.
With the addition of the multiplicative noise term $\epsilon^i_t$, the steady/equilibrium state, given by the fixed point $\mbox{\boldmath$\xi$}_0$,  is never exactly reached.
For low enough noise probability $p$, the system remains in the neighborhood of the fixed point associated with the expected pattern $\mbox{\boldmath$\xi$}_0$ but, as $p$ increases, the system can jump among other energy wells and, for high values of $p$, even visit all the configuration space, that is, all the possible stored patterns.
To take into account this temporal heterogeneity in the system dynamics, it is necessary to consider the time evolution of the overlap parameter, i.e., without going to the limit $t\rightarrow \infty$, that is:
\begin{equation}
Q_t = Q^0_t =  \frac{1}{N} \sum_{i=1}^N S^i_t \xi^i_0
\label{overlap_in_time}
\end{equation}
Then, we can compute time statistics of $Q_t$
and exploit these statistical indices as order parameters. In particular, we focus on two order parameters. The first one is given by the time average of $Q_t$:
\begin{equation}
\overline{Q} = \left\langle \frac{1}{N} \sum_{i=1}^N S^i_t \xi^i_0 \, \right\rangle
\label{overlap}
\end{equation}
being $\langle \cdot \rangle$ the time average carried out over the total simulation time $T_{sim}$.

\noindent
Regarding the second order parameter, we consider the second moment scaling $H$ of the diffusion process generated by the signal $Q_t$ \footnote{
  This is often denoted as Hurst exponent, even if it corresponds to the exponent introduced by Hurst in his milestone paper \cite{hurst_1951} only in the case of a monoscaling signal.
}. 
The diffusion variable $X_t$ is directly defined as the time integral of $Q_t$: 
$$
X_t = \int_0^t Q_s ds
$$ 
and the second moment scaling $H$ of $X_t$ is then evaluated by means of the Detrended Fluctuation Analysis (DFA) \cite{peng_pre94,peng_c1995} (see Appendix \ref{app:dfa}):
\beq
H = \lim_{t \rightarrow \infty} 
\frac {\ln G(t)}{\ln t}
\label{h_scaling_1}
\eeq
being $G(t)$ the DFA function defined by the following expression:
\begin{equation}
    \label{h_scaling_2}
    G^2(t) = \sigma^2(t) = \left\langle \left( X(t) - \overline{X}(t) \right)^2 \right\rangle \
\end{equation}
$\overline{X}(t)$ is a properly evaluated local trend of $X(t)$ (see Appendix \ref{app:dfa} for details). The hypothesis that $H$ is well-defined in equation (\ref{h_scaling_1}) is based on the assumption that $X_t$  is a self-similar, i.e., monoscaling signal, at least in the long-term regime, such that $X_t = a^H X_t$ and, consequently, $G(t) \sim t^{H}$ \cite{peng_c1995,stanley_pa1999,kantelhardt_pa2001}. 
The DFA is a tool of scaling (self-similarity) detection that is nowadays widely applied to characterize the complexity of time series (see, e.g., \cite{mallick_tac2021,wang_ieeesj2022}), revealing in particular the emergence of long-range temporal correlations and, thus, non-Markovian dynamics \cite{kantelhardt_pa2001,kantelhardt_pa2002,paradisi_npg12}, which are the signatures of emerging self-organizing behavior.
With respect to a direct evaluation of the second moment, the DFA
was proven to be robust with respect to the presence of spurious trends \cite{hu-stanley_pre2001} and is a widely used method to detect scaling in complex time series \cite{peng_pre94,peng_c1995}.

\noindent
In summary, $\overline{Q}$ and $H$ are the order parameters we are interested in. In particular,
$\overline{Q}$ is a one-time statistical average of the overlap parameter, while $H$ is related to two-time statistics of $Q_t$, and, in particular, on its temporal correlation \cite{peng_pre94}.

\section{Numerical Simulations and Results}
\label{sec:results}

\vspace{.5cm}
\noindent
{\bf Simulation setup}

\noindent
The MNIST dataset was utilized to train the neural network, storing
binary images with $784$ pixels ($28 \times 28$) \cite{deng2012mnist}. The size of MNIST images constrains the associative memory model to a number of neurons given by $N=784$.
MNIST is a dataset of digit images that is widely used as a benchmark and test case in machine learning studies and the digit images have both inter- and intra-patterns correlations. For this reason, we compare the MNIST with a dataset of standard Rademacher patterns. The Rademacher patterns are often exploited in theoretical studies of associative memory models, being characterized by total statistical independence both between and within patterns \cite{amit_pra1985,gardner_jpamg1987}.
In fact, Rademacher patterns are generated according to the Rademacher distribution \footnote{
This is essentially the same as the Bernoulli distribution, but on the set $\{-1,+1\}$ instead of $\{0,+1\}$.
}:
\beq
\xi^{R,i}_\mu \in \{-1; +1 \} \quad
Pr \{\xi^{R,i}_\mu = \pm 1\} = \frac12 \ ,
\label{rademacher}
\eeq
where $R$ stands for ``Rademacher''. As for the MNIST images, we build the Rademacher images with $N=784$ pixels.
Specifically, comprehensive simulations were conducted by varying the number $K$ of these binary images and
the noise probability $p$.
To assess how the network dynamics change based on the load $K$,
we used the following $K$ for the SEDAM model: $1$, $5$, $10$, $50$, from $100$ to $1000$ with incremental steps of $100$, and from $2000$ to $10000$ with incremental steps of $1000$. 
Conversely, for the SHN model that, as well-known, has a much lower storage capacity, we used the following $K$: $1$, $10$, $50$, $100$, $120$, $150$.
To investigate the effects of noise and identify potential critical regions, $25$ distinct $p$ values were evaluated ($p=0.001, 0.01, 0.1, 0.2$, from $0.21$ to $0.4$ with incremental steps of $0.01$, and $0.5$). 
For Rademacher-based models, to correctly assess the critical region, $18$ additional noise levels were included for the SHN model (from $0.0001$ to $0.0009$ in steps of $0.0001$, and from $p=0.11$ to $0.19$), and $9$ for the SEDAM (from $p=0.41$ to $0.49$).
To further refine this parametric analysis, three intermediate values were inserted between each previously identified critical onset point and the value immediately preceding it, resulting in a total of $43$ noise levels simulated for the MNIST-SEDAM, $90$ for the Rademacher-SHN, and $49$ for the Rademacher-SEDAM.

\noindent
The simulated networks consisted of $N=784$ neurons and were run for $T_{sim} = 200000$ time steps to ensure a thorough characterization of their temporal dynamics. 
%
%
The simulation time $T_{sim}$ was chosen to achieve the variation in the DFA slope (see next section ``Results''), which signals the emergence of anomalous diffusion, with a statistical time average that is sufficiently accurate for the DFA calculation itself (see Appendix \ref{app:dfa}). In all simulations, networks were initialized with a corrupted version of the first stored pattern ($\widetilde{\mbox{\boldmath$\xi$}}_0$), with $10 \%$ of pixels randomly flipped.

\vspace{1cm}
\noindent
{\bf Results}
 
\begin{figure}[h!]
    \includegraphics[width=0.49\linewidth]{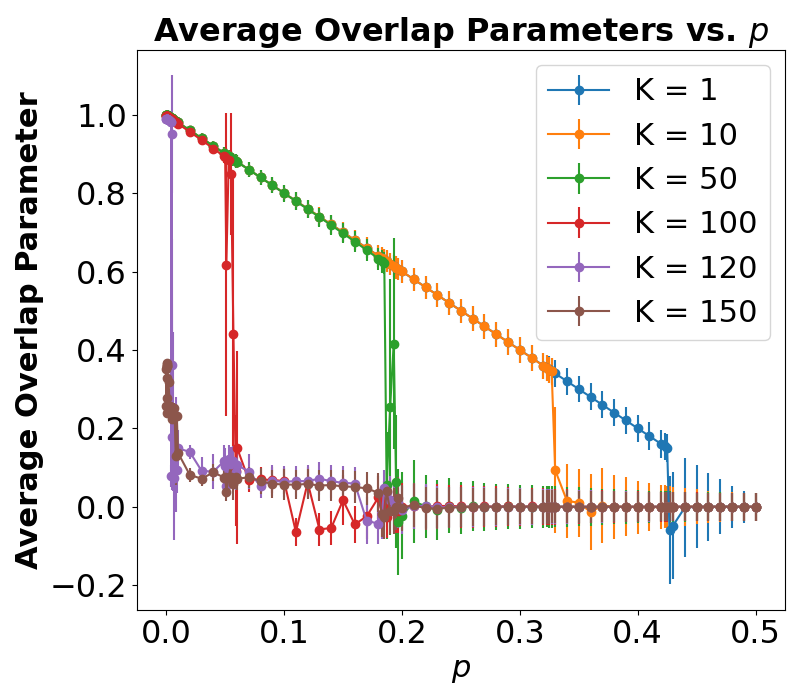}
    \includegraphics[width=0.49\linewidth]{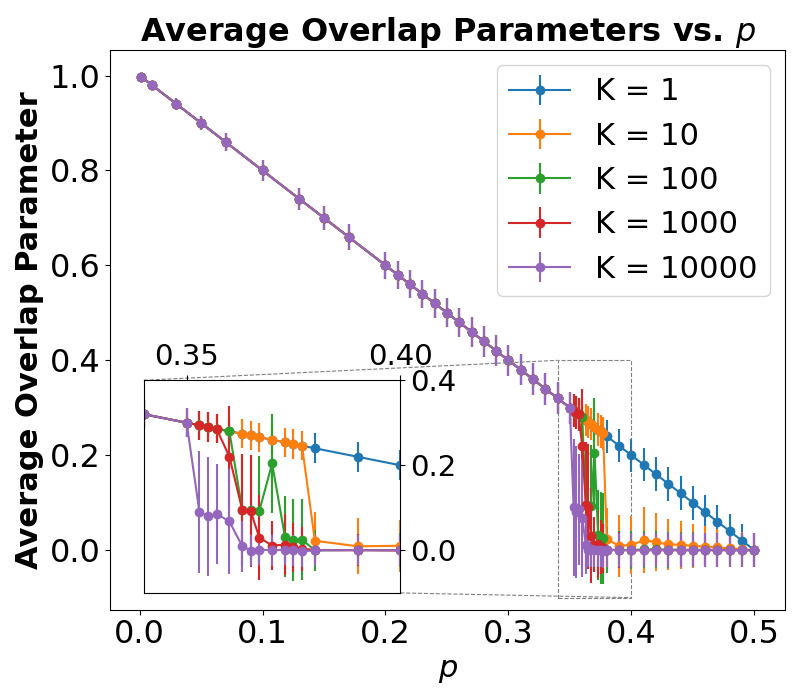}
    \includegraphics[width=0.49\linewidth]{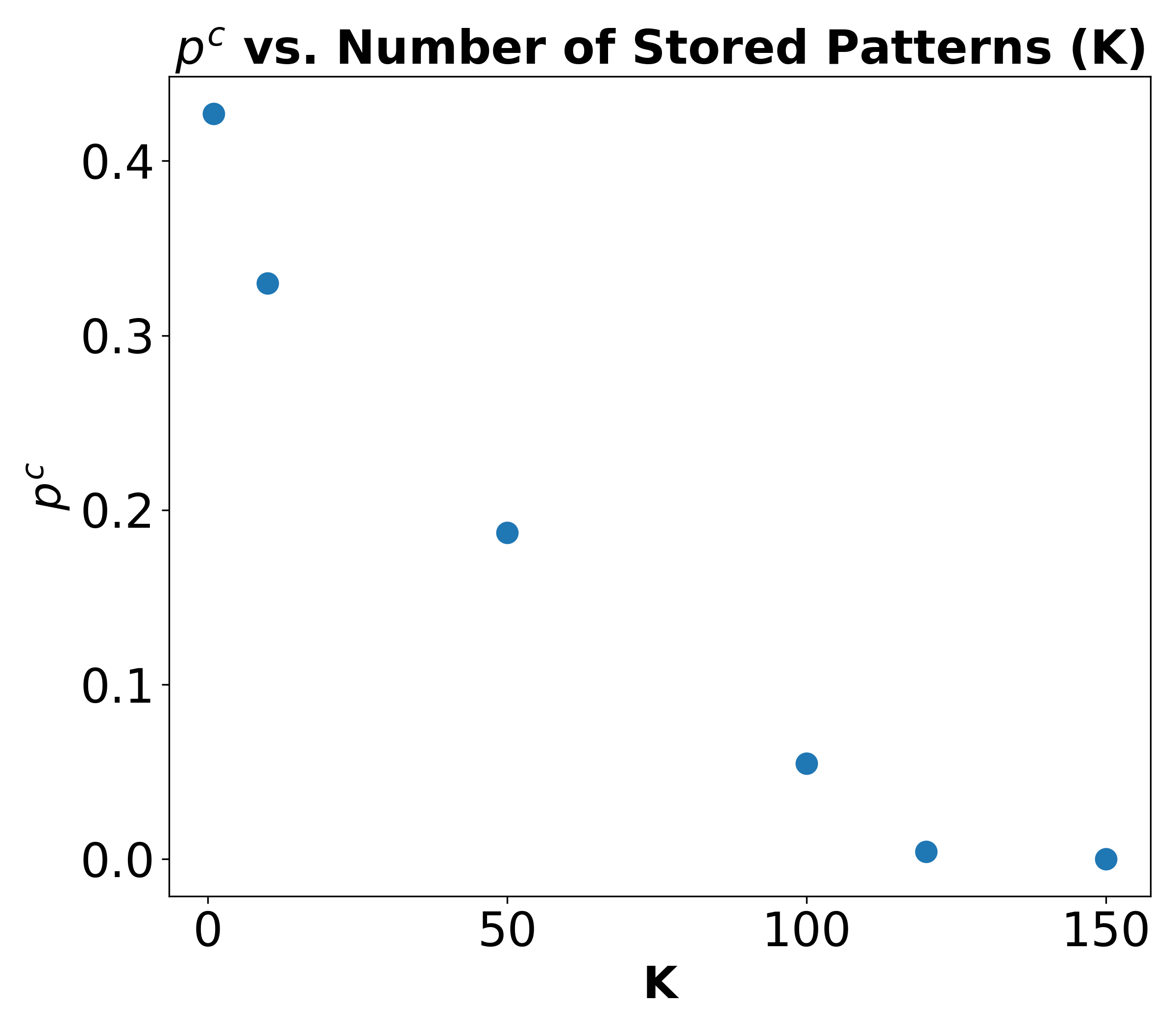}
    \includegraphics[width=0.49\linewidth]{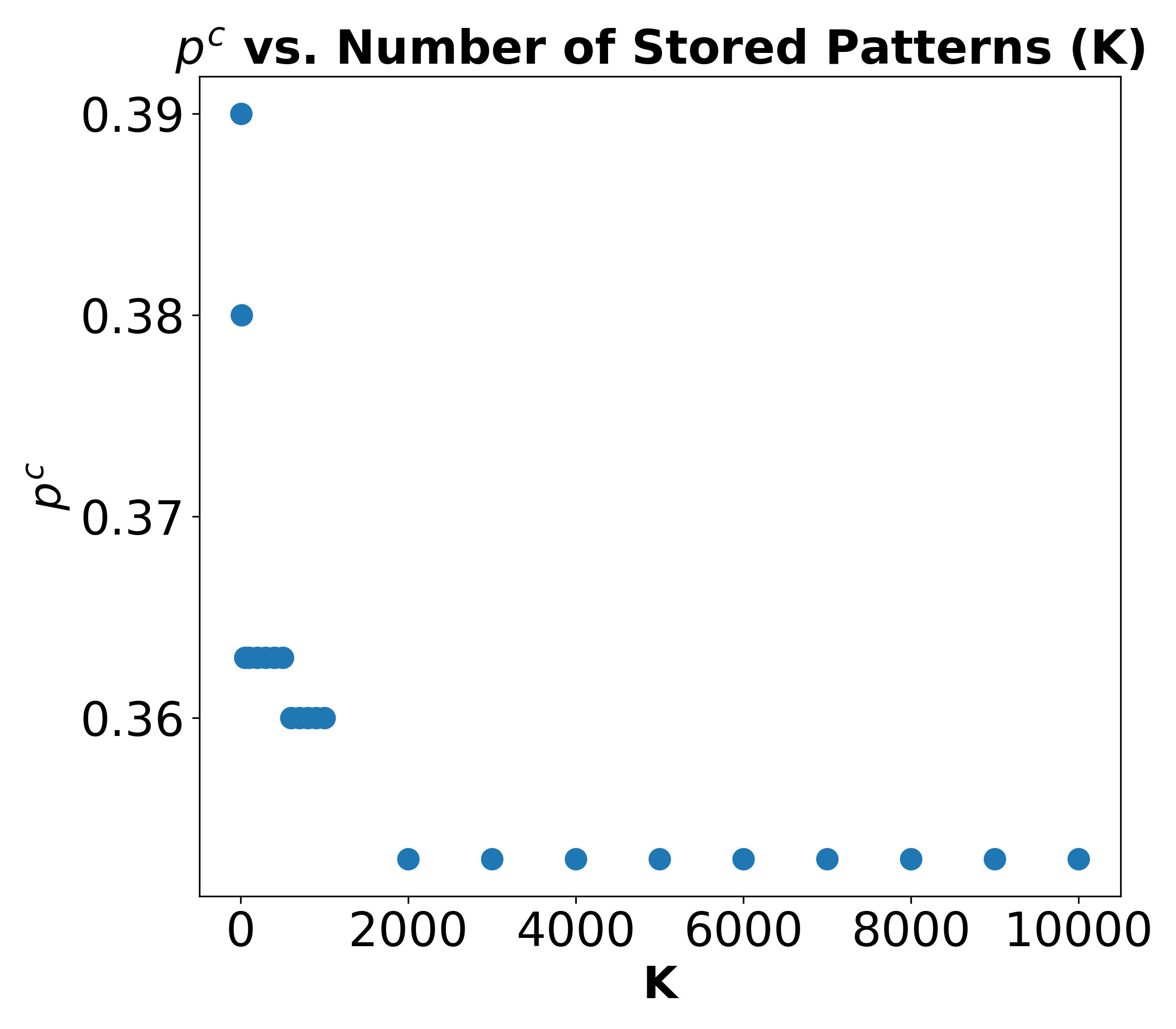}
    \caption{Results for stored Rademacher patterns. Top panels: average overlap parameter ${\overline Q}$ against $p$ by increasing the number $K$ of stored Rademacher patterns. Bottom panels: Onset critical noise levels ($p^c$) against the number $K$ of stored Rademacher patterns. Left panels: stochastic $2$-body HN model (SHN). Right panels: SEDAM model.}
    \label{fig:HN_Rademacher_overlap}
\end{figure}

\begin{figure}[h!] 
    \includegraphics[width=0.49\linewidth]{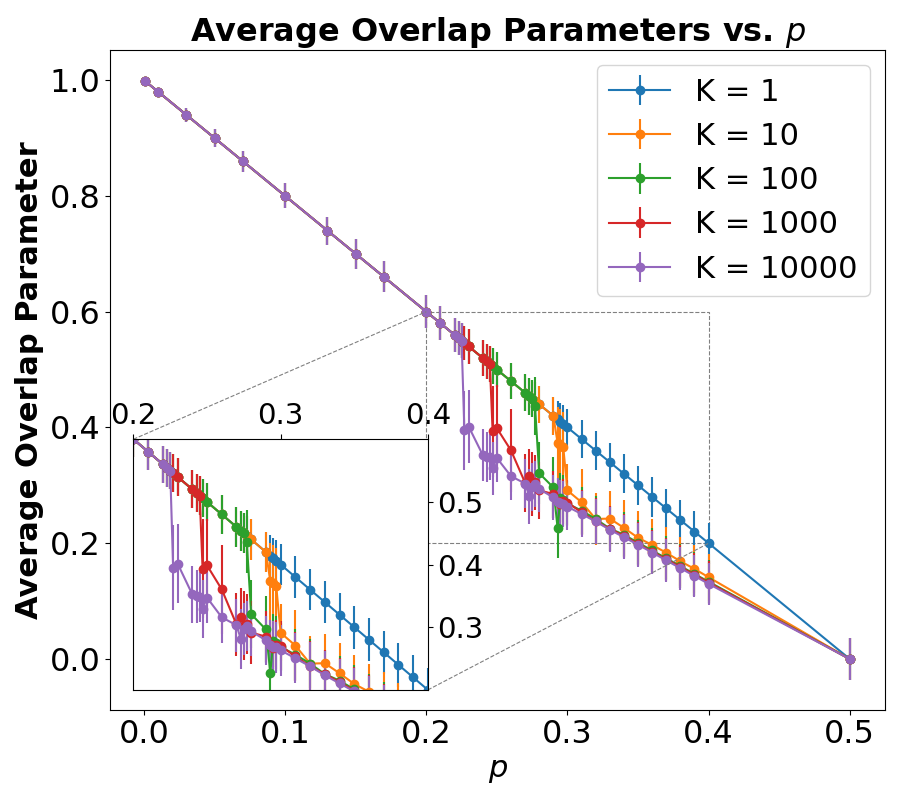}
    \includegraphics[width=0.49\linewidth]{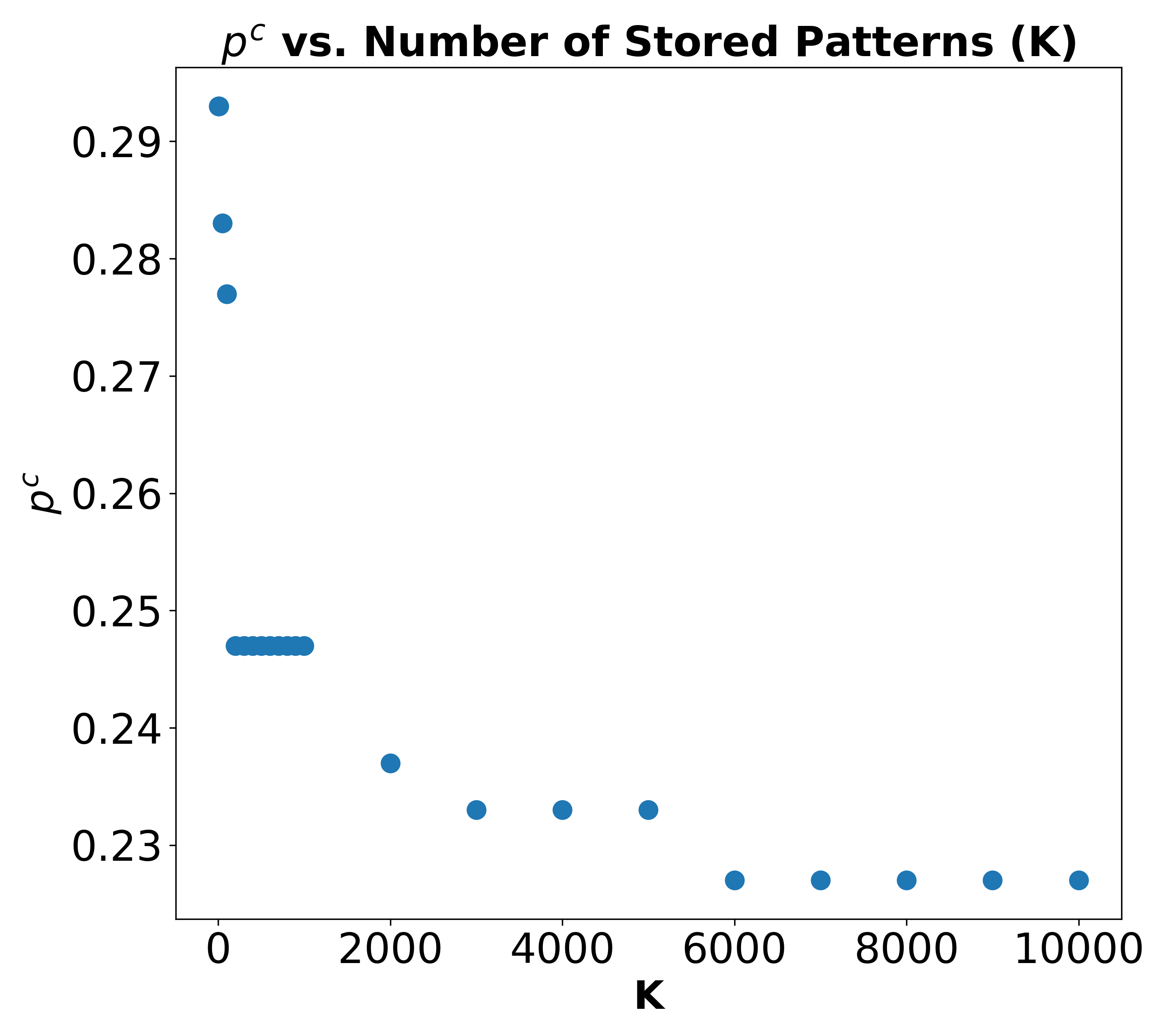}
    \caption{Results for the SEDAM model with MNIST stored patterns. Left panel: average overlap parameter ${\overline Q}$ against $p$ by increasing the number $K$ of stored patterns. Right panel: critical noise levels ($p^c$) against the number $K$ of MNIST stored patterns.}   \label{fig:SEMHN_MNIST_overlap_parameter}
\end{figure}

\noindent
Fig. \ref{fig:HN_Rademacher_overlap} shows the results of numerical simulations on both SHN and SEDAM models with stored Rademacher patterns. The average overlap parameters $\overline{Q}$ are reported and compared in the top panels. Both models exhibit a distinct phase transition resembling a second-order phase transition, as highlighted by the inset plot, with an almost vertical yet gradual drop of $\overline{Q}$, which depends on the noise level $p$, with the transition point shifting systematically as the number $K$ of stored patterns increases. 
In particular, the critical onset \footnote{
By ``onset'' value, we mean the first value $p^c$ when a scaling $H \ne 0.5$ is evaluated starting from the sub-critical phase.
} noise threshold ($p^c$) 
decreases progressively as the pattern capacity increases. Interestingly, the SEDAM model shows a critical drop to zero in $\overline{Q}$ at significantly higher noise intensities. This is also seen in the bottom panels, where the inverse relationship between the critical onset noise level $p^c$  and the load $K$ is illustrated. For the SHN, the noise intensity level at which the retrieval capacity breaks down
is about $p^c = 0.0001$ already at $K=150$ patterns, which is in line with the theoretical estimation of the maximum pattern load $K_c = \alpha_cN\simeq 0.14N \simeq 110$. At the same time, the SEDAM has a drop in the retrieval efficiency for noise levels $p^c \simeq 0.353$ even for $K=10000$ patterns. 
Actually, the SEDAM has a relatively fast decline from $p^c \simeq 0.39$ to $p^c \simeq 0.353$ as $K$ goes from $1$ to about $2000$ and remains constant up to $K=10000$ stored Rademacher patterns.
Interestingly, the range of critical noise values for the SEDAM model lies in a very narrow range, i.e., $p^c \sim 0.353-0.39$, while the range for the SHN model is much broader, i.e., $p^c \sim 0.0001-0.427$.\\
In Fig. \ref{fig:SEMHN_MNIST_overlap_parameter}, we report the results of numerical simulations for the SEDAM model with stored MNIST patterns. 
Conversely, the SHN model with MNIST patterns shows very low values of $\overline{Q}$ even for $K=3$ patterns and very low or zero noise intensity values, and, for this reason, this case is not reported in the plots.
A notable observation is that, 
for all models, the critical phase transition is not seen at a single critical point.
In particular, the critical transition of the MNIST-based SEDAM model occurs within a relatively narrow noise range, although wider than that of the counterpart Rademacher-based SEDAM model, specifically in the range  $p \sim 0.227-0.293$. 
Another interesting observation is that $\overline Q$ in the MNIST-based model (left panel of Fig. \ref{fig:SEMHN_MNIST_overlap_parameter}) does not drop to zero as in the Rademacher-based models, but to a different slope with respect to the noise level $p$, thus maintaining a non-zero degree of overlap, albeit a reduced one.
%
%
Further, the relationship between pattern storage capacity and noise tolerance is nonlinear: rather than following a simple linear decline, the critical noise level ($p^c$) displays a fast decrease
with increasing $K$. Moreover, for the MNIST-based SEDAM model, the most pronounced decrease in critical noise level occurs in the low-capacity regime,
specifically between $K=100$ and $K=200$ where $p^c$ drops from $0.277$ to $0.247$). Beyond this initial steep decline, the decay rate becomes substantially slower, with $p^c$ reaching only $0.227$ at $K=10000$. Interestingly, it is easy to see that there are wide intervals of $K$ in which $p^c$ remains almost constant.
%
%
%

\begin{figure*}[t]
    \includegraphics[width=1\linewidth]{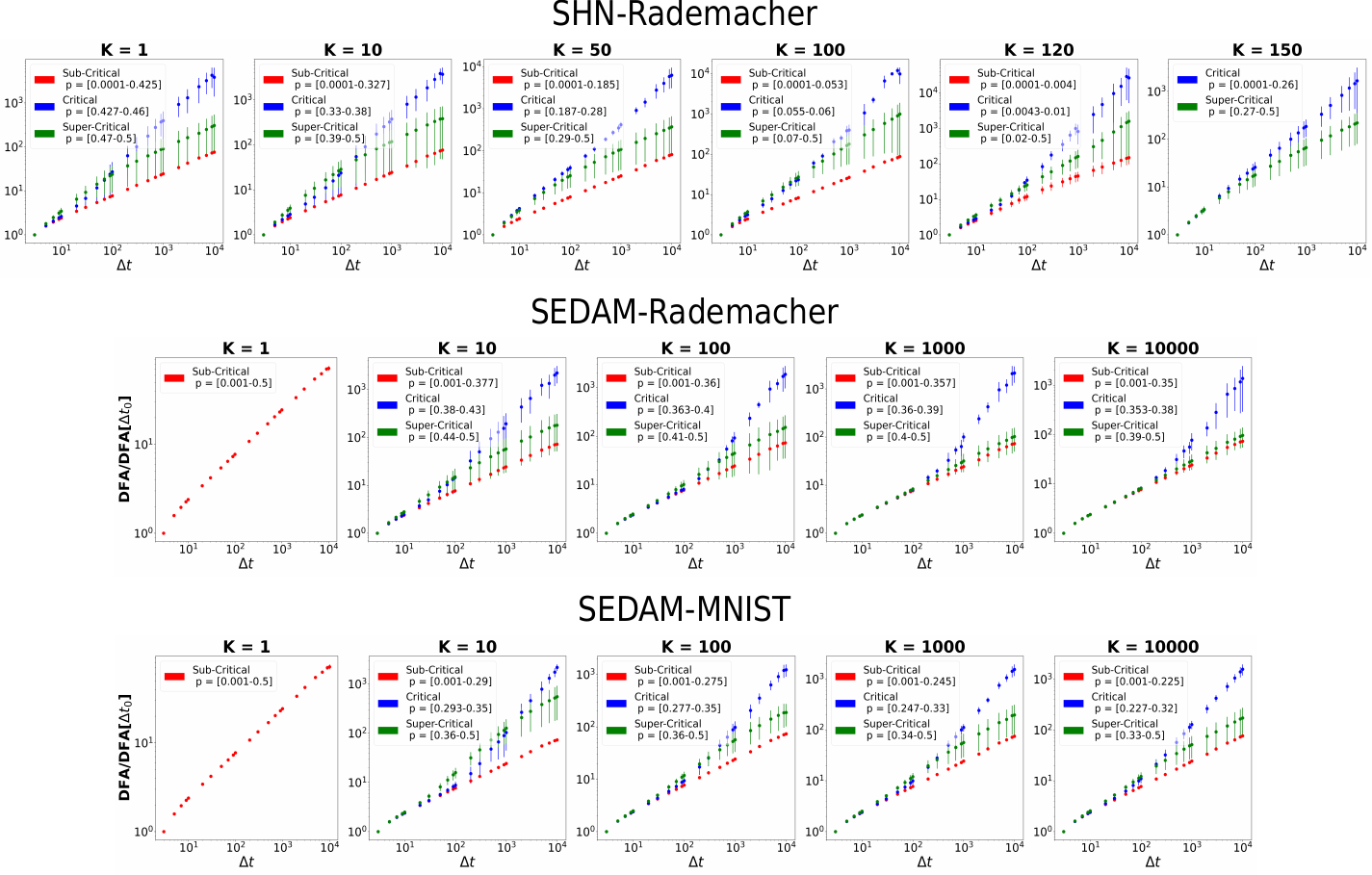}
    \caption{DFA curves computed across the three dynamical phases (sub-critical, critical, and super-critical) for the SHN-Rademacher, SEDAM-Rademacher, and SEDAM-MNIST models. Curves are normalized by $DFA(\Delta t_0)$, where $\Delta t_0$ denotes the smallest time lag considered. Each curve represents the mean (points) $\pm$ standard deviation (vertical bars) averaged over the noise intensity values $p$ indicated in the legend.}
    \label{fig:DFA_different_N}
\end{figure*}

\begin{figure}[h!]
\centering
\includegraphics[width=0.49\linewidth]{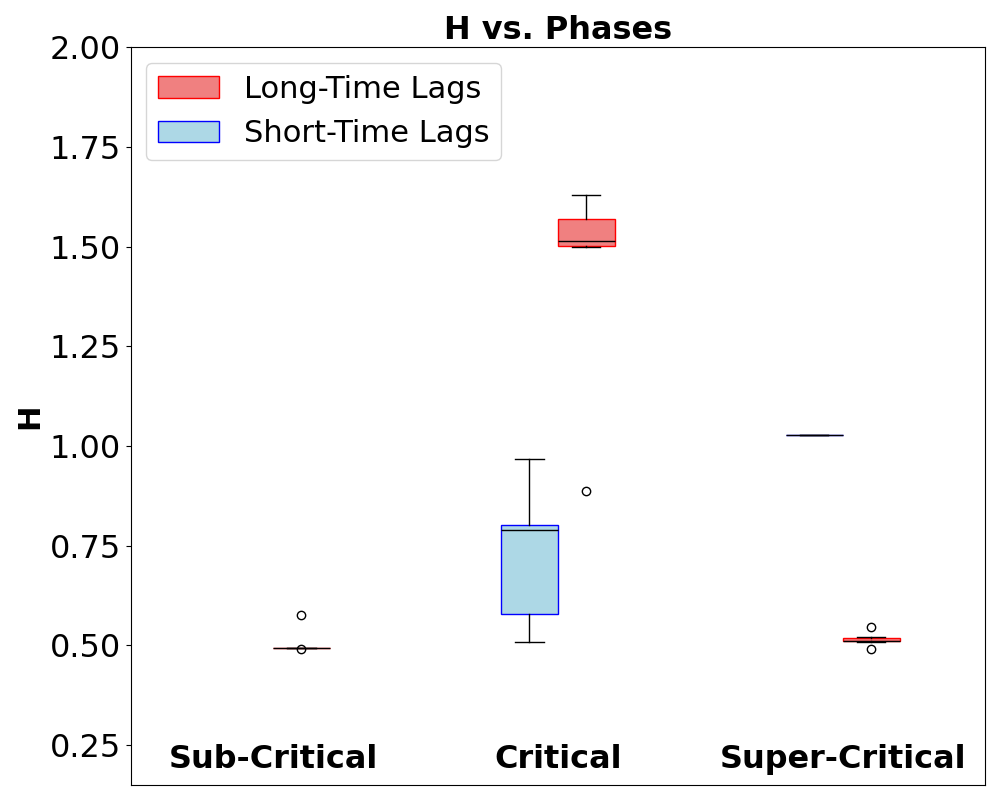}
\includegraphics[width=0.49\linewidth]{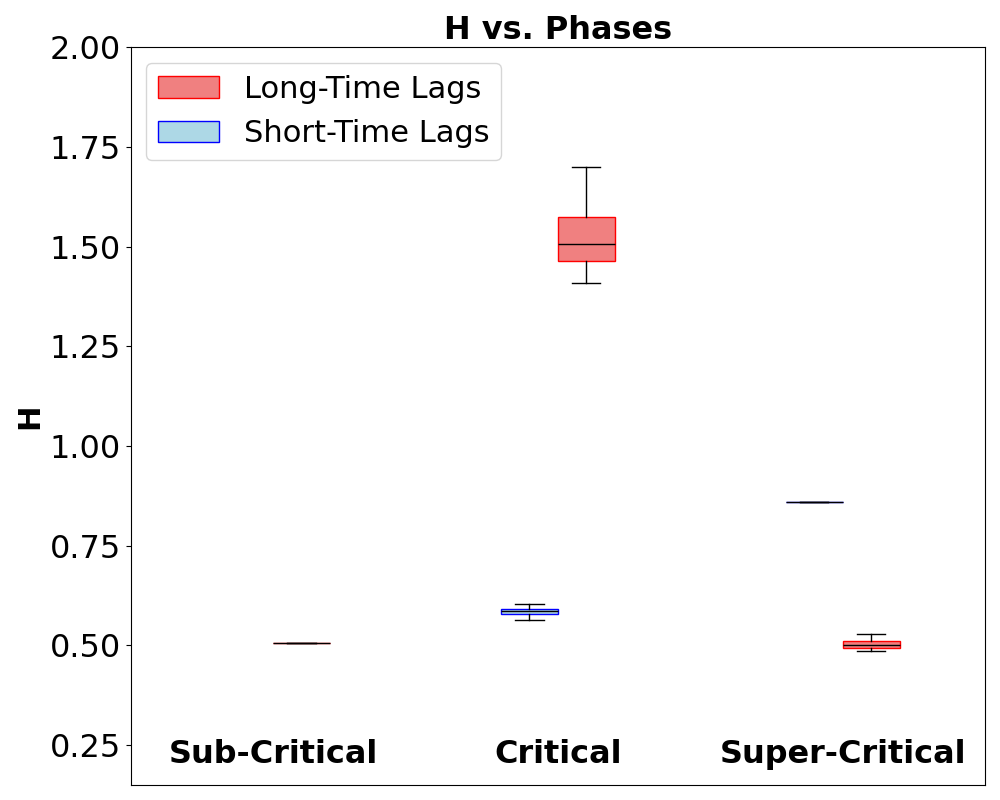}
\includegraphics[width=0.49\linewidth]{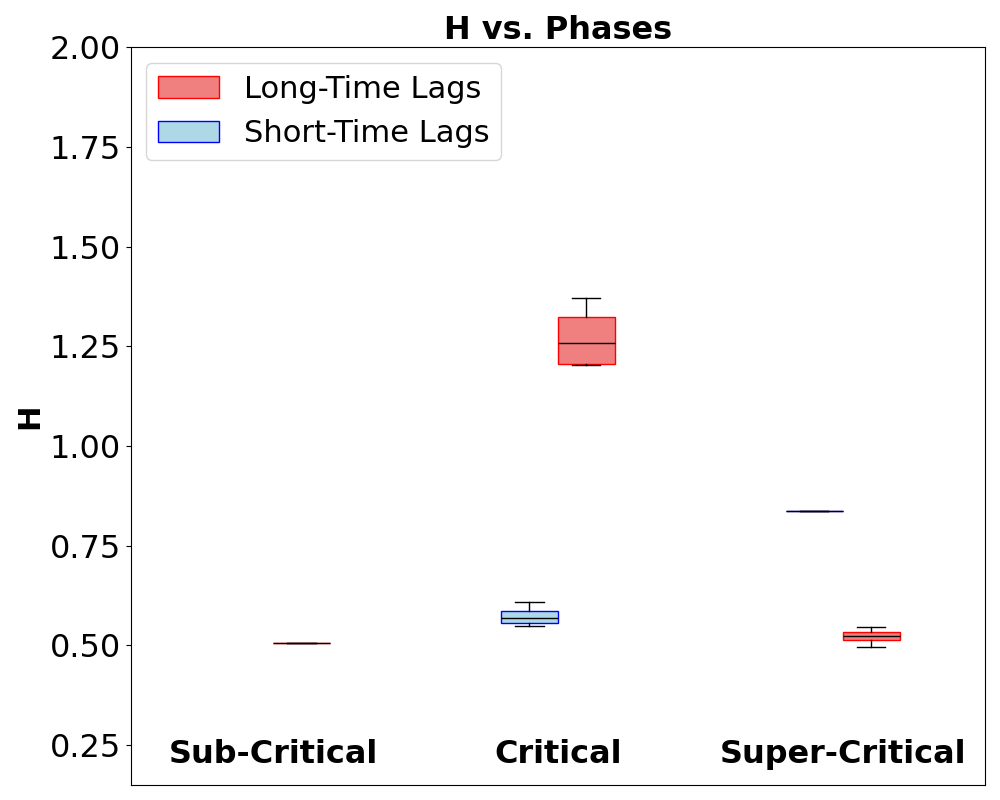}
\caption{Statistical boxplots of second moment scaling $H$ estimated from DFA in the three different phases (sub-critical, critical, super-critical) for both short and long time lag regimes. \\
Left Top Panel: boxplots of $H$ values for 2-body SHN model with Rademacher patterns. Right Top Panel: boxplots of $H$ values for SEDAM model with Rademacher patterns. Bottom Panel: boxplots of $H$ values for SEDAM model with MNIST patterns.
Statistics of $H$ is made over all $K$, with the central part reporting only fitted $H$ values corresponding to critical onset $p^c$ values.
Sub-critical conditions (left) use $p=0.1$ for the top right and bottom panels, and $p=0.0001$ for the top left panel. Super-critical conditions (right) use $p=0.47$ (top left), $p=0.45$ (top right), and $p=0.4$ (bottom panel).
}
\label{fig:H_vs_phases}
\end{figure}

\noindent
The results of DFA analysis are shown in Figs. \ref{fig:DFA_different_N} and \ref{fig:H_vs_phases}, which highlight the presence of a critical region marked by the strong discontinuity in the scaling exponent $H$. The discontinuity points were found to depend on $K$ and start at the same critical onset values $p^c$ of the average overlap ${\overline Q}$. Specifically, the critical onset $p^c$ varies with the number of stored patterns $K$: it spans $p^c \simeq 0.0001$–$0.427$ for the Rademacher-based SHN model, $p^c \simeq 0.353$–$0.38$ for the Rademacher-based SEDAM model, and $p^c \sim 0.227$–$0.293$ for the MNIST-based SEDAM model.

\noindent
Fig. \ref{fig:DFA_different_N} displays and compares the DFA curves across the sub-critical, critical and super-critical phases for all models studied and for representative values of $K$. 
%
A net change in the power index $H$ of the DFA function ($G(t) \sim t^H$) is evident when comparing the blue curves (obtained at criticality) with the red (sub-critical) and green (super-critical) curves.
Further, a crossover between short-time and long-time scaling exponents $H$ always emerges in the critical region. 
In some cases, this crossover also occurs in the super-critical phase—particularly at low values of $K$—but the long-range exponent is significantly smaller than the critical one, consistently taking the value $H \sim 0.5$ (see Fig. \ref{fig:H_vs_phases}).

\noindent
Notably, for the Rademacher-SHN model a critical region already appears at $K=1$. This result  indicates that the classical Hopfield model is less robust to noise than dense associative models, in which the scaling exponent remains unchanged as noise increases for the same $K$. Furthermore, at $K=150$ the Rademacher-SHN model exhibits no sub-critical region for any noise level. This demonstrates that when an associative model exceeds its maximum storage capacity, it does not transition directly to random dynamics—characterized by arbitrary jumps between potential wells—but instead passes through a critical region characterized by strong temporal correlations before entering the super-critical regime at higher noise.

\noindent
For the SEDAM models, three distinct dynamical regions remain identifiable even at very high storage capacity ($K=10000$). An important role is played by the intrinsic structure of the stored patterns. When uncorrelated patterns (Rademacher, middle panel of Fig. \ref{fig:DFA_different_N}) are stored, the sub- and super-critical regions exhibit very similar scaling behaviours, both differing markedly from the critical region even at small $K$. By contrast, when highly correlated patterns (MNIST, bottom panel of Fig. \ref{fig:DFA_different_N}) are stored, the super-critical and critical regions share overlapping of scaling behaviour on a more extended range of short-time scales-
an overlap that displays a net decrease as $K$ increases. This confirms that storing highly correlated patterns makes the SEDAM model less stable than storing intrinsically uncorrelated ones.

\noindent
Fig. \ref{fig:H_vs_phases} provides a synthetic overview of the application of DFA to numerically simulated data for the Rademacher-based SHN (top left panel) and SEDAM models (top right panel) and for the MNIST-based SEDAM model (bottom panel).
The central part of each panel
reports the statistics of the best fit estimations of $H$ for all $K$ and, for each $K$, we consider only the onset critical noise level $p^c$.
%
%
Both short-time and long-time best-fit estimations of DFA index $H$ are reported. 
Firstly, we note that, for all models, in the sub-critical regime, there is always a unique (long-time) scaling exponent $H \simeq 0.5$ and no short-time $H$ emerge. On the contrary, the behaviour changes in the critical and super-critical regimes depending on the model (SHN or SEDAM) and on the dataset (Rademacher or MNIST).
We observe that the Rademacher-based SHN model displays a higher short-time scaling exponent in the critical region ($H \sim 0.75$) compared to the SEDAM models, where the short-time exponent approaches $H \sim 0.5$.
It is worth noting that, 
the values $H>0.5$ can be seen in the short-time range of the super-critical region, but only for few cases, specifically, in the selected noise values, $K=10$ for the SEDAM model and $K = 1$ for the SHN model, while $H\simeq0.5$ is found in all other cases.\\
%
%
However, the most interesting behaviour emerges in the long-time $H$ exponent for all the considered models, where a clear phase transition is observed.
In particular, the transition is marked by a sharp increase from $H \approx 0.5$, i.e., standard Brownian diffusion,
in the sub-critical and super-critical phases, 
to a large anomalous super-diffusion scaling $H>1$, which is a signature of strongly persistent long-range correlations. 
Slight differences can be seen in the range of critical long-time $H$ values. In particular, between the Rademacher-based and the MNIST-based SEDAM models, the long-time $H$ exponent passes from a median $H \simeq 1.5$ to $H \simeq 1.25$, a discrepancy that can be attributed to the presence of inter- and/or intra-pattern correlations in the MNIST dataset.

%
%
%
%
%
%
    
\section{Discussion} 
\label{sec:discuss}

\noindent
In all the investigated models, the dynamical interplay between the
pattern storing 
and the multiplicative noise $\epsilon^i_{t} $, is the basic mechanism giving rise to the phase transitions emerging in both the average overlap parameter and the diffusion scaling $H$, 
this last one characterizing the temporal memory of the dynamics driving the time-dependent
overlap parameter itself.
%
Our findings are then consistent with the statistical physics of associative memory networks, which predicts that storage capacity and noise tolerance are fundamentally coupled \cite{Tsodyks_1988,Shiino_1992,theriault_spp2024}. 
However, in this work we applied a somewhat different approach while remaining consistent with principles and tools borrowed from statistical physics.
The novelties of the present approach to the study of criticality in DAM models basically are: \\
(i) the focus on out-of-equilibrium temporal dynamics that we get by introducing a stochastic version of the network models with a multiplicative noise, also avoiding any equilibrium assumption; \\
(ii) the application of DFA, which allowed to investigate the emergence of long-range time correlations in both SEDAM and SHN dynamical models;\\
(iii) the exploitation of the MNIST dataset, whose patterns are both inter- and intra-correlated. 

\noindent
Regarding point (i), the multiplicative noise $\epsilon_{t}^i$ in the SEDAM model of Eq. (\ref{modern_hop_exp_noise}) and in the SHN model of Eq. (\ref{classical_hop_noise}) 
resembles the effect of distortion errors in a dynamical image retrieval process. This dynamical modeling approach is then of interest not only in computer science, but also in the biology of perception, involving both {\it in vivo}, {\it in vitro} and {\it in silico} studies and is essentially different from the thermal noise introduced through the canonical ensemble approach that, moreover, cannot account for out-of-equilibrium effects.

\noindent
The application of DFA (point (ii)) to the temporal dynamics of the investigated models allowed us 
to discover that the phase transition is characterized not only by the overlap parameter, but also by a dynamic feature of the network involving temporal correlations (the diffusion scaling $H$). 
In fact, in correspondence of the ${\overline Q}$
phase transition, the $H$ displays an abrupt change from $0.5$ in both sub- and super-critical regions, signaling short-term memory and Markovian dynamics, to a strong super-diffusive regime $H>1$, thus marking the emergence of persistent long-range temporal correlations and, thus, non-Markovian dynamics in the critical region.
%
This emergence of strong persistent long-range time correlations with $H>1$
could be the signature of emerging self-organization, an aspect deserving future investigations.\\
Concerning point (iii), it is worth noting that most theoretical works, including most recent ones,  assume patterns without intra- and inter-pattern correlations, e.g, through the typical use of Rademacher patterns \cite{gardner_jpamg1987} and only a few works are devoted to the effect of intra- \cite{demarzo_pa2023,negri_prl2023} or inter-pattern correlations \cite{amit_pra1987,gardner_epl1987}. These studies are often limited to a specific kind of correlations, and only a few works exploited MNIST in the context of HN and DAM models \cite{krotov_2016modernHopf_first,musa2025denseassociativememorynonlinear}.
%
%
The comparison with the case of standard Rademacher patterns allowed us to highlight the effect of intra- and inter-pattern correlations, which characterize the MNIST dataset \cite{deng2012mnist} and, by construction, are absent in the Rademacher patterns. The SEDAM model has proven to have higher noise tolerance with Rademacher than with MNIST patterns, which is not a foregone conclusion but was expected.
On the contrary, the discrepancy between the temporal memory features of Rademacher-based ($H\sim 1.5$) and MNIST-based ($H\sim 1.25$) SEDAM models is not trivial and is a direct consequence of the MNIST pattern correlations.\\
%
%
%

\section{Concluding remarks}
\label{sec:concl}

\noindent
The central finding of this work is that, while the sub- and super-critical dynamics are Markovian and Gaussian, strongly persistent long-range temporal correlations emerge in the critical transition region. 
In the sub-critical case, this comes from the uncorrelated fluctuations around the correct minimum (almost fully retrieval). 
The super-critical case is less trivial and our analysis highlights the presence of short-term correlations that, over long times, resemble 
a Markovian random walk over a lattice, being the lattice sites given by the energy wells.
Conversely, the critical phase transition was here shown to be compatible with the emergence of a non-trivial critical state over a interval of noise intensity values that is not reduced to a single point \cite{bailly-longo_book2011,longo_fp2012}.
Looking at the time-averaged overlap $\overline{Q}$ vs. $p$, this small transition region is clearly compatible with a critical second-order phase transition. In the same region, the diffusion scaling $H$ displays values significantly greater than one, thus signaling a strong super-diffusion regime compatible with pronounced persistent long-range correlations and non-Markovian dynamics.
This long-range {\it ``temporal memory''} \footnote{
The term ``temporal memory'' is here specifically introduced to distinguish this feature from the ``associative memory'' feature of HN and DAM models, which was introduced since the 1970s to actually mean the ability of these same models to associate between input patterns and stored patterns. 
} 
characterizes, in the critical region, the persistence of stored patterns in the dynamics of the SEDAM model. This is a crucial difference with respect to canonical thermodynamical approach to DAM models (see, e.g., \cite{amit_pra1985,amit_prl1985,amit_ap1987,amit_book2005_critical_statphys,gardner_epl1987,demircigil2017modernHopfExp,theriault_spp2024,albanese_pa2026}), 
as the retrieval operation is here explicitly modeled as a dynamical process, given by the SEDAM and the SHN models, and not as a thermodynamical system in the equilibrium canonical ensemble assumption.
%

\noindent
In summary, the sub- and super-critical regions of the SEDAM model display normal diffusion and Brownian motion, thus implying Markovian dynamics of fluctuations in the network, while in the critical region, 
whose extent and memory features depend on the particular dataset (Rademacher or MNIST) considered and on the number of patterns $K$, the anomalous diffusion scaling exponent $H$ indicates that the dynamics are characterized by persistent long-range, i.e., non-Markovian, memory.
Similar observations can also be made regarding the SHN model, though it is important to note that its storage capacity and noise tolerance are significantly lower.
\section*{Acknowledgments}

\noindent
This work was supported by the Next-Generation-EU programme under the funding schemes PNRR-PE-AI scheme (M4C2, investment 1.3, line on AI)
FAIR ``Future Artificial Intelligence Research'', grant id PE00000013, Spoke-8: Pervasive AI.

\noindent
We would like to thank Laura Sebastiani for useful discussions.

\section*{Author Declarations}

\noindent
{\bf Conflict of interest}\\
The authors have no conflicts of interest to disclose.

\noindent
{\bf Author contributions}\\
Both authors are main authors and contributed equally to the work with the same roles, except: MC: Software, Visualization and PP: Supervision, Funding Acquisition, Project administration.

\section*{Data availability}

\noindent
MNIST data were used for the numerical simulations and can be found on the public free repository (see Ref. \cite{deng2012mnist}).

\appendix



\section{Detrended Fluctuation Analysis}
\label{app:dfa}

\noindent
Detrended Fluctuation Analysis (DFA) is a widely used method to detect scaling in time series and was proven to be robust with respect to the presence of spurious trends \cite{peng_pre94}. Being related to the self-similarity and long-range correlations of non-stationary time series, DFA is nowadays a widely recognized tool used in the field of complex self-organizing systems \cite{stanley_pa1999, kantelhardt_pa2001, paradisi_npg12, paradisi_springer2017, nayak_jhe2018}.

\noindent
The DFA algorithm starts by considering a signal $x_t$\footnote{
In the present work, $x_t$ is the overlap parameter $Q_t$.
}.
In the first step of the algorithm, the global average $\langle x \rangle$ of the signal is computed and then subtracted from the signal itself. After that, the ``diffusion variable" $X(t)$ is computed:
\begin{equation}
    X(t) = \sum_{t} (x_t - \langle x \rangle)
\end{equation}
The obtained signal is divided into non-overlapping windows of length $\Delta t$: $[k\Delta t + 1, k\Delta t + \Delta t]$. For each time window, the local trend is evaluated with a least-squares straight-line fit:
\begin{equation}
    \label{local_trend_DFA}
    \overline{X}_{k, \Delta t}(t) = a_{k, \Delta t} t + b_{k, \Delta t}; \quad k \Delta t < t \leq (k + 1) \Delta t 
\end{equation}
Subsequently, the fluctuation is evaluated as:
\begin{equation}
    \label{fluctuation_DFA}
    \Tilde{X}_{k, \Delta t}(t) = X(t) - \overline{X}_{k, \Delta t}(t); \quad k \Delta t < t \leq (k + 1) \Delta t
\end{equation}
Over every time window $\Delta t$ the mean square deviation is evaluated as:
\begin{equation}
\label{mean_square_deviation_fluctuation}
    G^2(k, \Delta t) = \frac{1}{\Delta t} \sum_{t = k \Delta t +1}^{(k + 1) \Delta t} \Tilde{X}^2_{k, \Delta t} (t)
\end{equation}
Finally, an average over the windows is performed:
\begin{equation}
\label{average_of_mean_square_deviations}
    G^2(\Delta t) = \frac{1}{[M/\Delta t]} \sum_{k=0}^{[M/\Delta t]} F^2(k, \Delta t)
\end{equation}
To assess if the process is self-similar the relation $G(\Delta t) \sim \Delta t^H$ must be true. To evaluate that the following relation is considered:
\begin{equation}
    \label{dfa_function}
    \log (G(\Delta t)) = H \log (\Delta t) + C
\end{equation}
Therefore, a simple linear fit procedure on a logarithmic scale is sufficient to estimate the value of $H$. \\
To carry out the DFA calculations, we employed the function MFDFA of the package MFDFA of Python \cite{gorjao2022MFDFA}. \\
\ \\
%

\section{Synthetic list of notations, symbols and acronyms}
\label{app:notations}

\noindent
$N$: dimension of neural network (number of neurons)\\
$K$: number of stored patterns\\
$
\mbox{\boldmath$\xi$}_\mu=[\xi^1_\mu,\xi^2_\mu,...,\xi^N_\mu]$: $\mu$-th stored pattern ($1 \le \mu \le K$)\\
${\bf S}_t = [S^1_t,S^2_t,...,S^N_t]$: state of the network at time $t$ ($S^i_t$ = state of $i$-th neuron at time $t$)\\
$\mbox{\boldmath$\xi$}^T$: transpose of vector ${\bf \xi}$\\
$\mbox{\boldmath$\xi$}^T \mbox{\boldmath$\eta$}$: scalar product of vectors $\mbox{\boldmath$\xi$}$ and $\mbox{\boldmath$\eta$}$\\
$n$: power exponent of the $n$-body interaction function $F(z) = z^n$\\
$\epsilon^i_{t}$: salt-and-pepper noise of the $i$-th neuron with values in $\{-1,+1\}$\\
HN: Hopfield Network\\
SHN: Stochastic HN\\
MHN: Modern Hopfield Network\\
DAM: Dense Associative Memory\\
SEDAM: Stochastic Exponential DAM



\bibliography{Bibliography}






\end{document}